\documentclass[
    reprint,
    superscriptaddress,
    amsmath,amssymb,
    aps,
    prx,
    longbibliography
]{revtex4-1}

\usepackage{graphicx}
\usepackage{amsmath}
\usepackage{amssymb}
\usepackage{algorithm}
\usepackage[noend]{algpseudocode}

\usepackage[colorlinks=true,urlcolor=blue,linkcolor=blue,citecolor=blue]{hyperref}
\usepackage[margin=0.75in]{geometry}
\algrenewcommand\algorithmicrequire{\textbf{Input:}}
\algrenewcommand\algorithmicensure{\textbf{Output:}}
\usepackage{xcolor}

\begin{document}
\title{Large Scale Distributed Linear Algebra With Tensor Processing Units}

\author{Adam G.M. Lewis}
\affiliation{Sandbox@Alphabet, Mountain View, CA 94043, USA}

\author{Jackson Beall}
\affiliation{Sandbox@Alphabet, Mountain View, CA 94043, USA}

\author{Martin Ganahl}
\affiliation{Sandbox@Alphabet, Mountain View, CA 94043, USA}

\author{Markus Hauru}
\affiliation{Sandbox@Alphabet, Mountain View, CA 94043, USA}

\author{Shrestha Basu Mallick}
\affiliation{Sandbox@Alphabet, Mountain View, CA 94043, USA}

\author{Guifre Vidal}
\affiliation{Sandbox@Alphabet, Mountain View, CA 94043, USA}

\date{\today}

\begin{abstract}
We have repurposed Google Tensor Processing Units (TPUs), application-specific chips developed for machine learning, into large-scale dense linear algebra supercomputers. The TPUs' fast inter-core interconnects (ICI)s, physically two-dimensional network topology, and high-bandwidth memory (HBM) permit distributed matrix multiplication algorithms to rapidly become computationally bound. In this regime, the matrix-multiply units (MXU)s dominate the runtime, yielding impressive scaling, performance, and raw size: operating in float32 precision, a full 2048-core pod of third generation TPUs can multiply two matrices with linear size $N=2^{20} = 1\,048\,576$ in about 2 minutes. Via curated algorithms emphasizing large, single-core matrix multiplications, other tasks in dense linear algebra can similarly scale. As examples, we present (i) QR decomposition; (ii) resolution of linear systems; and (iii) the computation of matrix functions by polynomial iteration, demonstrated by the matrix polar factorization.
\end{abstract}

\maketitle

\section{Introduction}
Neural network inference and training requires low-precision multiplication of large matrices. To service this need, Google has reincarnated the systolic array as the Tensor Processing Unit (TPU). As is typical of ASICs, compared to CPUs at fixed wattage TPUs sacrifice flexibility for speed: they essentially only multiply matrices, but are very good at doing so. We have measured distributed, single precision (floating point 32 or fp32) matrix multiplication performance of around 21 PFLOPS --- competitive with an academic cluster allocation, while much more accessible and carbon-friendly --- on a third-generation TPU ``pod" of 2048 cores. But a means to harness such performance for scientific simulation is presently lacking.

We are aware of two approaches to such a harness, which we view as complementary. The first \cite{LearningPDE, LearningAptamer, LearningCFD, KohnShamRegularizer} 
accelerates some CPU-based computation with some kind of TPU-based machine learning algorithm, for example by using a neural network to precondition GMRES. The second, to which this paper contributes, curates traditional scientific algorithms to run efficiently on TPUs directly. Previous work by others in this vein has concerned discrete \cite{TPUFFT2} and fast \cite{TPUFFT1} distributed Fourier transforms, Monte-Carlo simulation \cite{TPUMonteCarlo}, and image processing \cite{huot2019highresolution}. Our group's sister papers address quantum circuit simulation \cite{tpu_circuit,tpu_Z2field}, many-body quantum physics \cite{tpu_qphys,tpu_floquet,tpu_qhardware}, electronic structure computation via density functional theory (DFT) \cite{tpu_qchem} and coupled cluster (CC) methods \cite{tpu_CC}, and tensor network algorithms such as the density matrix renormalization group (DMRG) \cite{tpu_DMRG}. 

\begin{figure*}
    \centering
    \includegraphics[scale=0.37]{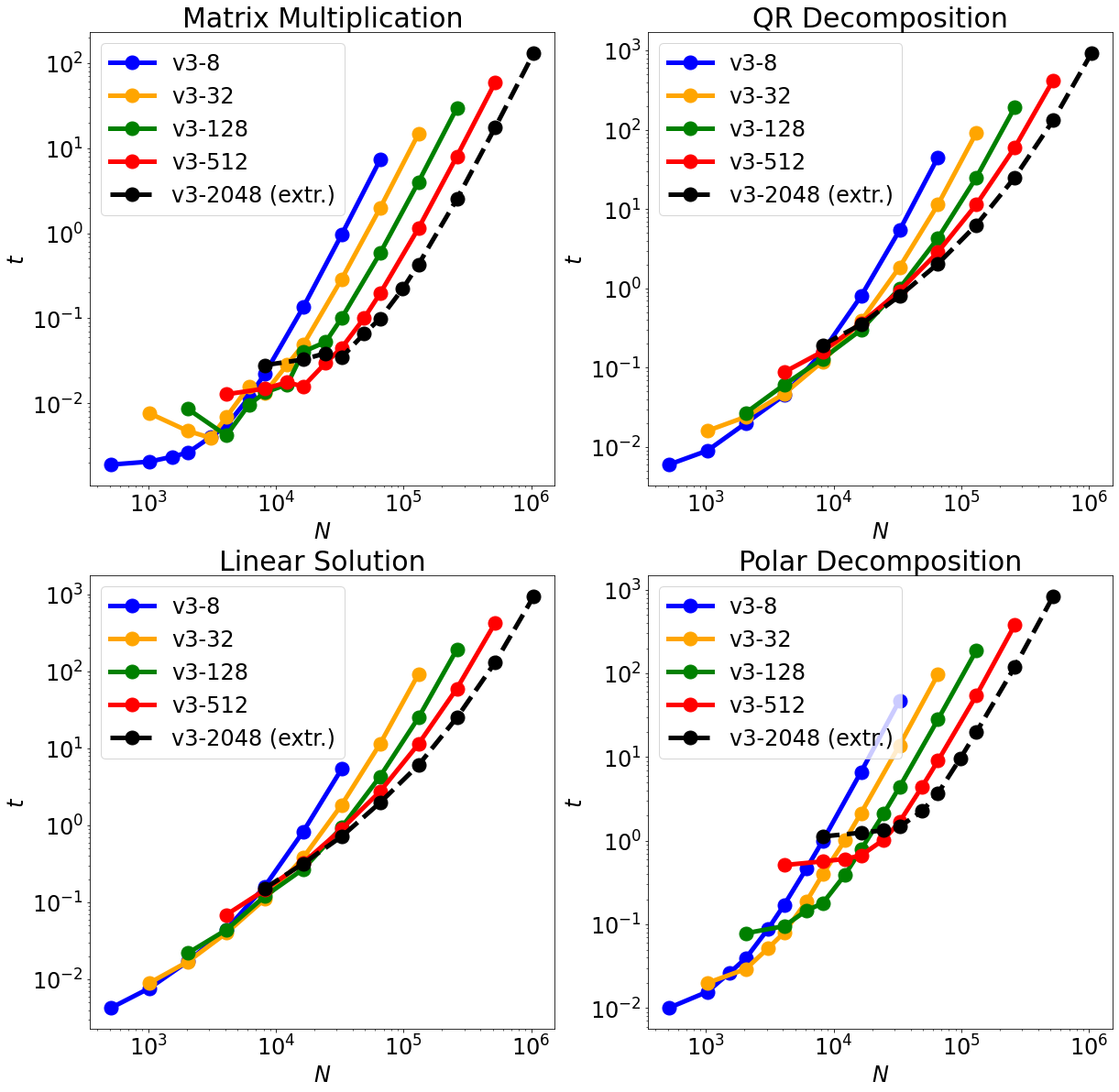}
    \caption{
Wallclock time $t$ in seconds vs linear size $N$ of square input for 
Top Left: Matrix Multiplication;
Top Right: QR Decomposition; 
Bottom Left: Linear Solution; 
Bottom Right: Polar Decomposition. 
The notation, e.g. v3-8, refers to a third generation (v3) TPU network of 8 cores (v3-8). The v3-2048 results are a linear extrapolation from v3-512, necessitated by temporary resource constraints.
As concrete examples, on a full TPUv3 pod (2048 cores), working with dense $(N,N)$ matrices of linear size $N=2^{20} = 1\, 048\, 576$, a matrix multiplication takes about 2 minutes, whereas both a QR decomposition and solving a linear system take about 20 minutes; similarly, the polar decomposition takes about 20 minutes for linear size $N = 2^{19}= 524\, 288$. 
    }
    \label{fig:PerformanceFigure}
\end{figure*}

This paper concerns the more foundational tasks of distributed dense linear algebra. While a single TPU core can already store and operate on large matrices (e.g. of size $(16\, 384, 32\, 768)$ in single precision\footnote{We consider TPU network topologies with a 2:1 aspect ratio, so that local blocks of square matrices have a corresponding 1:2 aspect ratio.}), the main advantage of TPUs is their ability to scale to full pods, which can handle much larger matrices (e.g. $(1\, 048\, 576, 1\, 048\, 576)$, or $2048\times$ larger size). Accordingly, our focus is in understanding how to perform distributed, multi-core versions of linear algebra operations whose single-core version is already provided by the JAX library. Specifically, in this paper we demonstrate four distributed dense linear algebra tasks at scale (see Figure \ref{fig:PerformanceFigure} for benchmarks):
 \renewcommand{\labelenumi}{\Alph{enumi})}
\begin{enumerate}
    \item Distributed matrix-multiplication, using the SUMMA \cite{SUMMA} algorithm to translate from the TPUs' efficient single-core matrix multiplication to comparably-efficient, distributed matrix multiplication without data replication.
    
    \item Distributed QR decomposition, using an adapted CAQR algorithm \cite{tsqr} emphasizing matrix multiplication.
    
    \item Solution of linear systems, implemented as a distributed QR decomposition followed by a distributed triangular solve.
    
    \item Distributed computation of matrix functions. As a specific example we show the polar decomposition, expressing a given matrix as the product of one unitary and one positive-semidefinite factor. 
\end{enumerate}

In sister papers we use some of these tasks to accelerate and scale-up a number of applications. For instance, a variant of SUMMA is used for CC computations \cite{tpu_CC}, the distributed QR decomposition is used for DMRG \cite{tpu_DMRG}, and distributed matrix functions similar to a polar decomposition, as well as the inverse square root, are used for the purification step of DFT \cite{tpu_qchem}. 

Two remarks are in order. The results of this paper refer exclusively to single precision. However, TPUs can also perform linear algebra in (emulated) double precision, as needed e.g. in some quantum chemistry applications \cite{tpu_qchem}. For matrix multiplications, this incurs a roughly $11\times$ increase in computational cost. Moreover, we also note that while the benchmark results presented here were obtained with third generation TPUs (denoted TPUv3), fourth generation TPUs (denoted TPUv4) are already available. A TPUv4 pod (8\,192 cores) can handle matrices with linear size 2$\times$ larger than a TPUv3 pod. 

The remainder of the paper is structured as follows. Section \ref{TPUSection} briefly explicates the TPU architecture. Section \ref{Tasks} explains our approach to the four tasks listed above and presents benchmarks. Section \ref{Conclusion} gives a closing discussion.

\section{Tensor Processing Units (TPUs)}
\label{TPUSection}
Each TPUv3 chip has two cores, each equipped with two ``matrix multiply units" (MXUs) --- systolic-arrays capable of multiplying two $(128, 128)$ matrices in 128 cycles. The chips are connected to one another via relatively fast interconnects, in a two-dimensional toroidal network ranging from 4 chips (8 cores) to 1024 chips (2048 cores) total, with each group of 4 chips (8 cores) controlled by a separate host CPU. See \cite{TPUinfo} for many more details on the TPU architecture.

TPUs natively perform bf16-precision matrix multiplication with fp32 accumulation. That is, the TPU stores and sums data as fp32, but each individual matrix multiplication of a floating-point number is done in a specialized low-precision format called ``brain float 16" or bf16, comparable to fp16 but with slightly more range and slightly less precision. The TPU can still operate in fp32 precision, however, via an internal mixed-precision algorithm which incurs a roughly $6 \times$ penalty in compute time.

One generally programs for the TPU using XLA \cite{XLA}, an optimized graph compiler proprietary to Google. XLA translates from roughly C-like commands called HLOs to roughly assembly-like equivalents called LLOs. The HLOs themselves may be written directly, but are usually instead ``traced" from any of several higher-level languages. We used Jax \cite{Jax}, a NumPy like interface to XLA.

The TPU architecture and its access via XLA introduces several constraints:
\begin{itemize}
    \item Since XLA requires prior knowledge of memory boundaries, there is limited support for dynamical array shapes. All shapes must be computable from ``static" data available at compile time, with changes to static data incurring an expensive recompilation. This can complicate algorithms involving e.g. a shrinking block size.
    \item TPUs are optimized to perform large matrix multiplications. Thus, a relatively straightforward path to their efficient use is to find algorithms which also involve large matrix multiplications.
    \item TPUs store data in physically two-dimensional memory, with each ``row" able to store an 8 by 128 matrix panel. Matrices whose dimensions are not divisible by 8 or 128 respectively are in effect zero-padded up to the next-largest sizes which are.
\end{itemize}

The subsequent discussion showcases a selection of distributed dense linear algebra algorithms that function well despite these constraints. We have chosen these specific algorithms because of their widespread use in scientific computing and/or their pivotal role in several applications in \cite{tpu_circuit, tpu_Z2field, tpu_qphys, tpu_floquet, tpu_qhardware, tpu_qchem, tpu_CC, tpu_DMRG} that were mentioned in the introduction.

\section{Distributed Linear Algebra Benchmarks}
\label{Tasks}

We target three ``core" tasks representing essential computations of e.g. LAPACK:
\begin{itemize}
    \item Matrix Multiplication: Computation of $\mathbf{C}$ in $\mathbf{A} \mathbf{B} = \mathbf{C}$ given matrices $\mathbf{A}$ and $\mathbf{B}$.
    \item QR Factorization: Computation of $\mathbf{Q}$ with orthonormal columns and upper-triangular $\mathbf{R}$ in $\mathbf{A} = \mathbf{Q} \mathbf{R}$ given $\mathbf{A}$.
    \item Linear Solution: Computation of $\mathbf{x}$ in $\mathbf{A} \mathbf{x} = \mathbf{b}$ given $\mathbf{A}$ and $\mathbf{b}$.
\end{itemize}
We also illustrate another task, for which TPUs turn out to be especially suitable:
\begin{itemize}
    \item Matrix Functions: Computation of $f(\mathbf{A})$, given a matrix $\mathbf{A}$ and a function $f(x)$, where $f(\mathbf{A})$ is a matrix obtained from $\mathbf{A}$ by transforming by $f$ (depending on context) either its singular values or its eigenvalues. 
\end{itemize}    

We will illustrate matrix functions explicitly with the polar factorization, which can be construed as the case where $f$ is the signum function acting on singular values, and thus mapping positive singular values to $+1$, and with matrix inversion. 

\subsection{Distributed Matrix Multiplication}
The first step is to build large-scale matrix multiplication from fast local matrix multiplication and fast inter-chip communications over a 2D toroidal topology. This can be achieved using any of a variety of distributed matrix multiplication algorithms. We use SUMMA \cite{SUMMA} (Scalable Universal Matrix Multiplication Algorithm), whose memory footprint is tuneable, and which straightforwardly handles transposed matrix multiplication.

SUMMA requires matrices be distributed across processors as two-dimensional blocks. A group of $p$ TPU cores is first divided into a $(p_r, p_c)$ processor grid. An $(M, N)$ matrix is then divided into $(m = M / p_r, n = N / p_c)$ blocks, and each block assigned to exactly one processor. The assignment must be ``adapted" to the matrix, meaning:
\begin{itemize}
    \item{Traversing through $p_r$ or $m$ with $n$ and $p_c$ fixed, also traverses through $M$ with $N$ fixed (row-adapted), and}
    \item{Traversing through $p_c$ or $n$ with $m$ and $p_r$ fixed, also traverses along $N$ with $M$ fixed (column-adapted).} 
\end{itemize}

\begin{figure}
    \centering
    \includegraphics[scale=1.1]{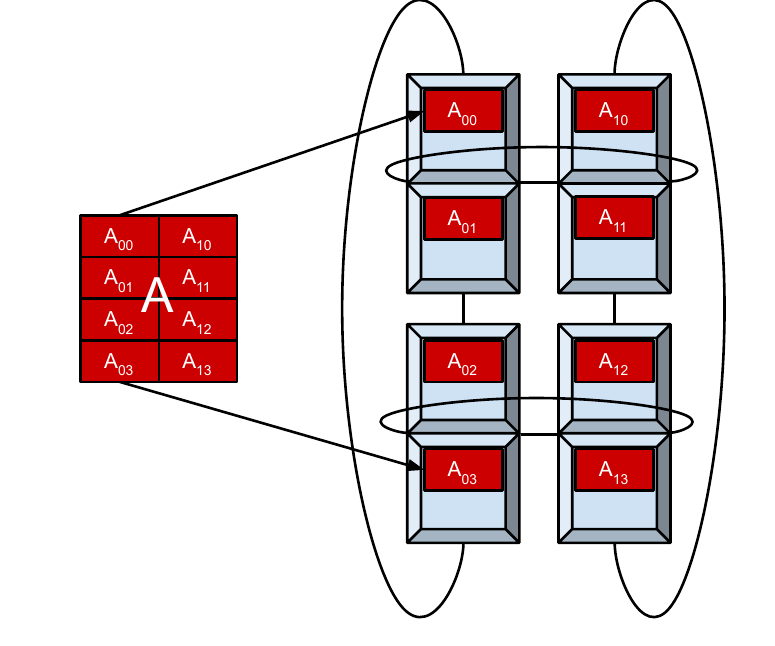}
    \caption{The matrix $\mathbf{A}$ is `checkerboard' distributed onto a (4, 2) ``v3-8" TPU grid, by partition into a corresponding (4, 2) grid of contiguous matrix blocks (red rectangles). The TPU cores are depicted as light blue squares, each separate chip as a pair of two adjacent such squares, and the 2D toroidal network connectivity between chips as black lines.}
    \label{fig:CheckerboardFig} 
\end{figure}

Though SUMMA does not require it, for simplicity we furthermore adopt the \textbf{checkerboard} distribution illustrated in Figure \ref{fig:CheckerboardFig}:
\begin{itemize}
    \item {$m$ and $p_r$ are contiguous in $M$, and} 
    \item {$n$ and $p_c$ are contiguous in $N$.}
\end{itemize}
We zero-pad as required when $p_r$ does not evenly divide $M$ or $p_c$ does not evenly divide $N$. Heuristically, the checkerboard distribution assigns matrix blocks to processors by overlaying the TPU grid (``checkerboard") atop the mathematical matrix.

\begin{figure}
    \centering
    \includegraphics[scale=0.3]{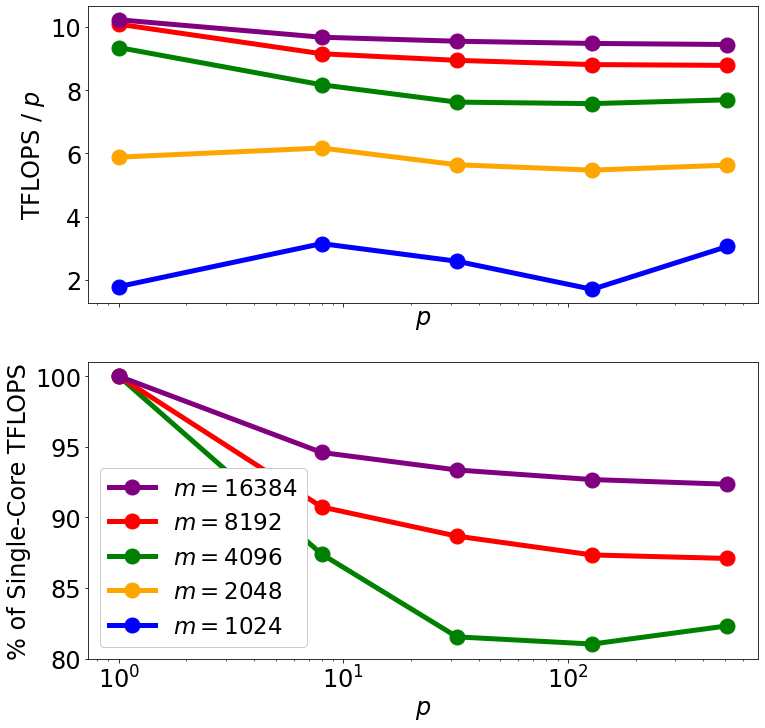}
    \caption{Weak scaling data for distributed TPU matrix multiplication. Each curve holds local matrix sizes, fixed by the number of rows $m$ per core, constant. The x-axis shows the number of TPU cores $p$. The top panel shows the TFLOPS \eqref{eq:TFLOPS} per core. The bottom shows what percentage of the corresponding $p=1$ value is attained by each point on the top three curves.}
    \label{fig:ScalingFigure}
\end{figure}

Distributed linear algebra packages more commonly adopt a \textbf{block cyclic} distribution, in which adjacent matrix blocks are assigned cyclically to adjacent processors, rather than contiguously in local memory as in the checkerboard distribution. This allows slices of the distributed matrix to be taken without affecting load balance. Sections \ref{QRSection} and \ref{LinearSection} will demonstrate algorithms which indeed suffer from the poor load balance of the checkerboard distribution. However, Figure \ref{fig:ScalingFigure} will also show that each TPU core must be fed a matrix of about $25$\% of the maximum available linear size to begin saturating the serial throughput of the MXUs. In practice, this need for very large block sizes makes the block cyclic distribution impractical.

\begin{figure}
    \centering
    \includegraphics[scale=1.1]{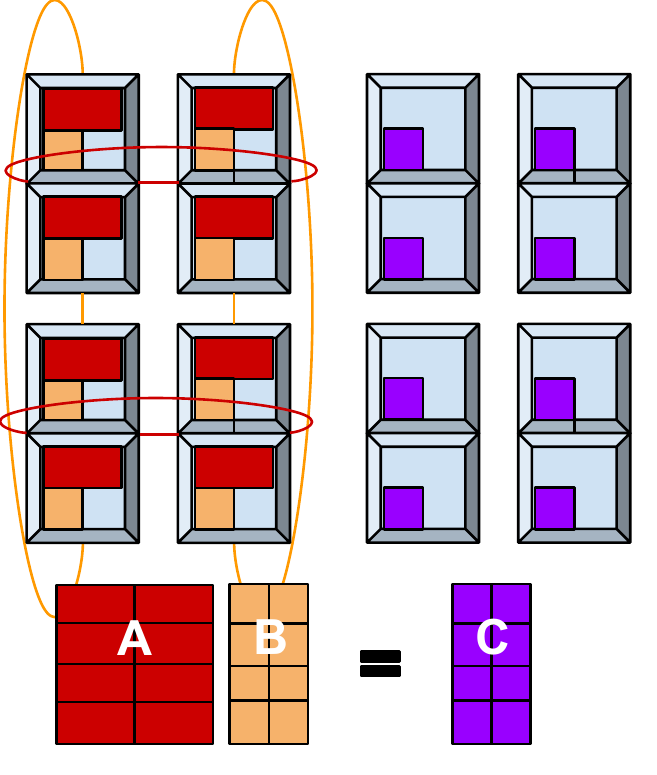}

    \caption{Matrix distributions before and after SUMMA matrix multiplication, $\mathbf{A} \mathbf{B} = \mathbf{C}$. Left: the matrix factors $\mathbf{A}$ (red) and $\mathbf{B}$ (orange) are checkerboard-distributed onto a v3-8 TPU grid. During multiplication, $\mathbf{A}$ will be communicated across processor columns (along the red-coloured interconnects) and $\mathbf{B}$ across processor rows (along the orange-coloured interconnects). Right: distribution of the result matrix $\mathbf{C}$.}
    \label{fig:SUMMAfig} 
\end{figure}

Now let us discuss the SUMMA algorithm. We will rehearse the untransposed case, and thus seek
\begin{equation}
    \label{eq:Matmul}
    C_{ij} = \sum^K_k A_{ik} B_{kj}.
\end{equation}
for an $(M, N)$ matrix $\mathbf{C}$, and $(M, K)$ matrix $\mathbf{A}$, and a $(K, N)$ matrix $\mathbf{B}$. It is convenient to also write the above equation as
\begin{equation}
    \label{eq:Matmul2}
    \mathbf{C} = \mathbf{A} \mathbf{B}.
\end{equation}

SUMMA works by dividing the $K$ values of index $k$ into $N_b$ ``panels" of $k_b$ entries each. We will use Greek letters to enumerate such panels, e.g. $\kappa$. Let us define corresponding matrix pannels $\mathbf{A}^{(\kappa)}$ and $\mathbf{B}^{(\kappa)}$ by 
\begin{equation}
    \label{eq:Panels}
    \mathbf{A}^{(\kappa)} \equiv A_{0:M-1~ k':k''} , ~~~~ \mathbf{B}^{(\kappa)} \equiv B_{k':k''~0:N-1}, 
    %
\end{equation}
where $k' = \kappa k_b$ and $k'' = (\kappa+1)k_b-1$.
Expressed in this notation, \eqref{eq:Matmul} becomes
\begin{equation}
    \label{eq:BlockMatmul}
    \mathbf{C} = \sum^{N_b}_\kappa \mathbf{A}^{(\kappa)}  \mathbf{B}^{(\kappa)}.
\end{equation}
Notice that each term in summand of Eq. \eqref{eq:BlockMatmul} is a matrix product, $\mathbf{C}^{(\kappa)} = \mathbf{A}^{(\kappa)}  \mathbf{B}^{(\kappa)}$. SUMMA works by paralellizing each individual such matrix product.

Given that $\mathbf{A}$ and $\mathbf{B}$ are already checkerboard distributed, the block column panel $\mathbf{A}^{(\kappa)}$ must therefore be broadcast to all other processor columns within processor rows, and the block row panel $\mathbf{B}^{(\kappa)}$ to all other processor rows within processor columns. Performing these broadcasts simultaneously exploits all four channels of each TPU chip in a pipelined fashion, with a maximum broadcasted distance of $\mathrm{max}(p_r, p_c) // 2$ (whether and how to pipeline in practice is decided automatically by the XLA compiler). The resulting matrix $\mathbf{C}$ inherits the same checkerboard distribution as the inputs, as illustrated in Figure \ref{fig:SUMMAfig}.

By choosing $k_b$ to be small relative to $m$ and $n$ but large enough to yield good single-core throughout (larger than about 512 in practice), this algorithm makes near-optimal use of TPU resources, while consuming negligible memory apart from that needed to store $\mathbf{A}$, $\mathbf{B}$, and $\mathbf{C}$. This is evinced in the top left panel of Figure \ref{fig:PerformanceFigure}, which shows the wallclock time required to multiply square fp32 matrices of size $N$ distributed across various TPU v3 configurations. 

The maximum value of $N$ is determined by the necessity to fit $\mathbf{A}$, $\mathbf{B}$, and $\mathbf{C}$ in memory, demonstrating SUMMA's negligible need for additional memory. For instance, on a full TPU pod (2048 cores) we can fit two ($N,N$) matrices of linear size $N = 2^{20} = 1\,048\,576$, which can then be multiplied in about 2 minutes. For large enough $N$, the straight lines on the log-log plot indicate runtime is dominated by $O(N^3)$ operations. 

The excellent scaling with increasing number of TPU cores $p$ can be seen by in turn consulting the two panels of Figure \ref{fig:ScalingFigure}. $p$ here is the x-axis, while each line holds the number of matrix rows per core $m = \frac{N}{\sqrt{2 p}}$ fixed. Notice the undistributed $p=1$ case, which does not invoke SUMMA, is also included. 

Figure \ref{fig:ScalingFigure} invokes the throughput speed of the operations in TFLOPS,
\begin{equation}
    \label{eq:TFLOPS}
    \mathrm{TFLOPS} \equiv \frac{2N^3}{t \cdot 10^{12}},
\end{equation}
where $t$ is the measured wallclock time in seconds. Very heuristically, the \eqref{eq:TFLOPS} measure the number of multiplications and additions implicitly performed by the TPUs per second. The top panel plots the TFLOPS per core ($\frac{\mathrm{TFLOPS}}{p}$) against $p$. We see the $p=1$ performance only begins to saturate (to a bit more than 10 TFLOPS) around $m=4096$, which is an appreciable fraction of the memory available per core. As alluded to earlier, this motivates our choice of a checkerboard rather than a block-cyclic distribution, since the latter would necessitate smaller local blocks and thus significantly degrade performance in all but the largest cases. 

Optimal scaling would be indicated by flat horizontal lines. For large $m$ we quite nearly reach this optimum, as depicted quantitatively in the bottom panel, which shows the percentage of the corresponding $p=1$ value attained by each point of the top three curves. For $m=16384$ this is quite nearly 95\%. The non-monotonicity of the bottom two curves is presumably a consequence of the operation not being fully computationally bound here.

In this study we are primarily concerned with the large $N$ regime. Scaling is less favourable for small $N$, both within and between TPU configurations. Two problems occur when $N$ is small: the block outer products in \eqref{eq:BlockMatmul} become too small to obtain good serial throughput from the TPU cores, and the constant overhead cost to initiate a communication becomes important relative to the cost of communication itself. Smaller $N$ performance could be improved, if needed, by exploiting the extra available memory. By copying $\mathbf{A}$, $\mathbf{B}$, and $\mathbf{C}$ between some or all processors rather than distributing among them, the individual summands in \eqref{eq:BlockMatmul} can be evaluated in parallel; this strategy is sometimes known as a ``2.5 D algorithm". Similar considerations could be applied to the QR and matrix function algorithms.

\subsection{QR Factorization}
\label{QRSection}

The QR factorization rewrites an $(M, N)$ matrix $\mathbf{A}$ with $M \ge N$ as the product of a ``Q-factor" with orthonormal columns and an upper-triangular ``R-factor", $\mathbf{A} = \mathbf{Q} \mathbf{R}$. Two closely related factorizations can be distinguished: the ``full" factorization, with $\mathbf{Q}$ $(M, M)$ and $\mathbf{R}$ $(M, N)$; and the ``reduced" factorization, with $\mathbf{Q}$ $(M, N)$ and $\mathbf{R}$ $(N, N)$. Both cases serve as a primitive in many applications, since for example the reduced $\mathbf{Q}$ factor orthonormally spans the column-space of $\mathbf{A}$.

\begin{figure}
    \centering
    \includegraphics[scale=0.3]{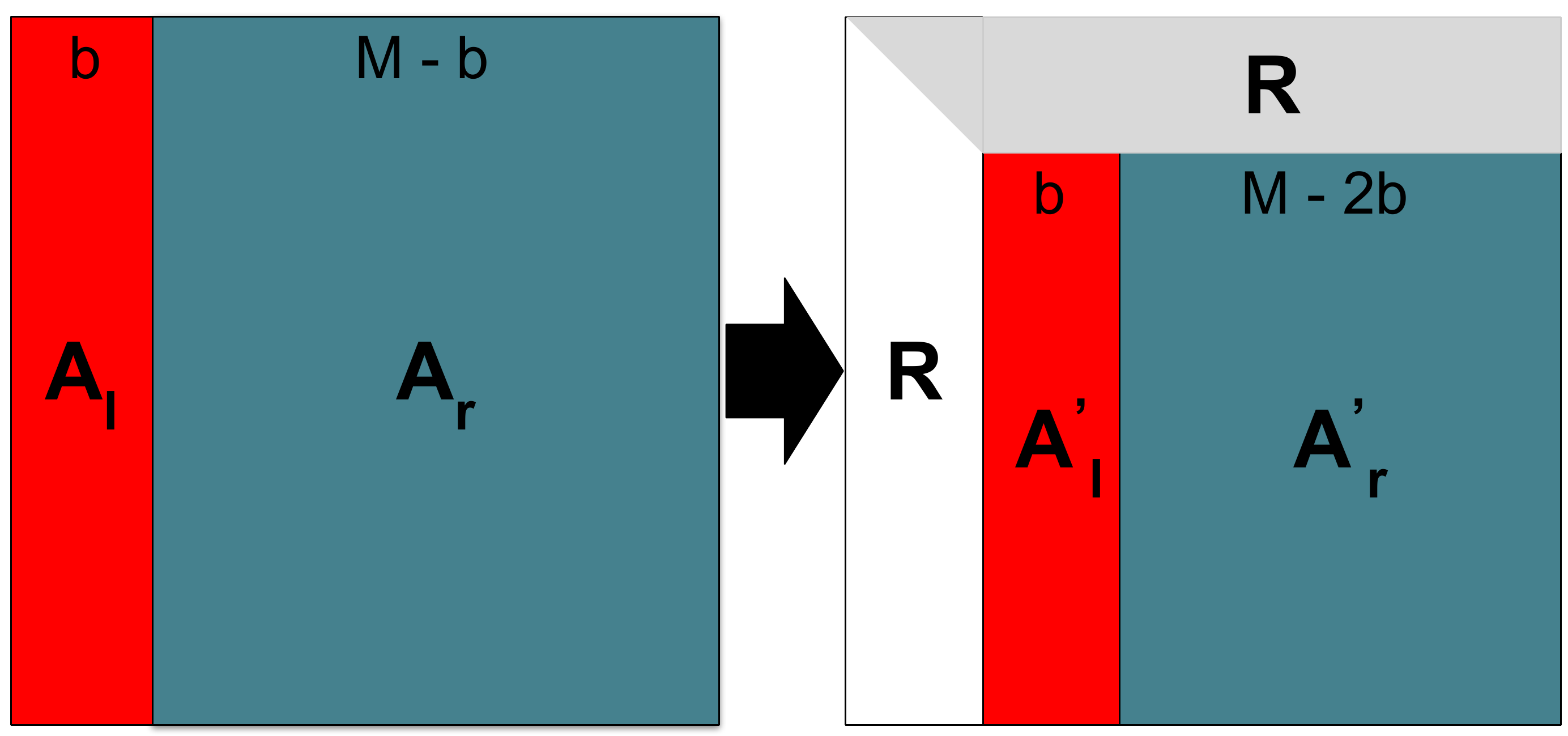}
    \caption{ Depiction of the CAQR algorithm \cite{tsqr} used to factor matrices distributed across two dimensional processor grids. Each iteration uses TSQR to factor the column panel $\mathbf{A}_\mathbf{l}$, first into its reduced Q factor, and then into an implicit WY representation of its full Q factor. The latter is then applied to the panel $\mathbf{A}_\mathbf{r}$, replacing a new strip of $\mathbf{A}$ with data from its $\mathbf{R}$ factor. The process is then repeated with new $\mathbf{A}_\mathbf{l}$ and $\mathbf{A}_\mathbf{r}$ (labelled $\mathbf{A}'_\mathbf{l}$ and $\mathbf{A}'_\mathbf{r}$, right of the arrow in the figure) until $\mathbf{R}$ has fully replaced $\mathbf{A}$.}
    \label{fig:CAQRFigure}
\end{figure}

Jax via XLA provides an efficient and stable single-core QR factorization algorithm based on blocked Householder transformations as described in \cite{MatrixComputations}. We focus here on distributing the computation over TPU grids, using a suitably adjusted version of the CAQR algorithm of \cite{tsqr}. In brief, our approach is as follows:
\begin{enumerate}
    \item A panel of $b$ columns of $\mathbf{A}$ is selected, labelled $\mathbf{A}_l$ in Figure \ref{fig:CAQRFigure}.
    \item \emph{Column factorization:} The full QR decomposition of that panel is implicitly computed, $\mathbf{A}_l = \mathbf{Q}_f \mathbf{R}_f$. $\mathbf{A}_l$ is replaced with $\mathbf{R}_f$.
    \item \emph{Panel update:} The remaining columns $\mathbf{A_r}$ are replaced by $\mathbf{Q}_f^H \mathbf{A}_r$.
\end{enumerate}
The above basic procedure is known as ``right-looking block QR". Typically, the \emph{column factorization} step would be handed by computing ``Householder" representations of individual columns of $\mathbf{Q}_f$ one by one, but this involves too many scalar operations on TPUs. 

\begin{figure}
    \centering
    \includegraphics[scale=0.8]{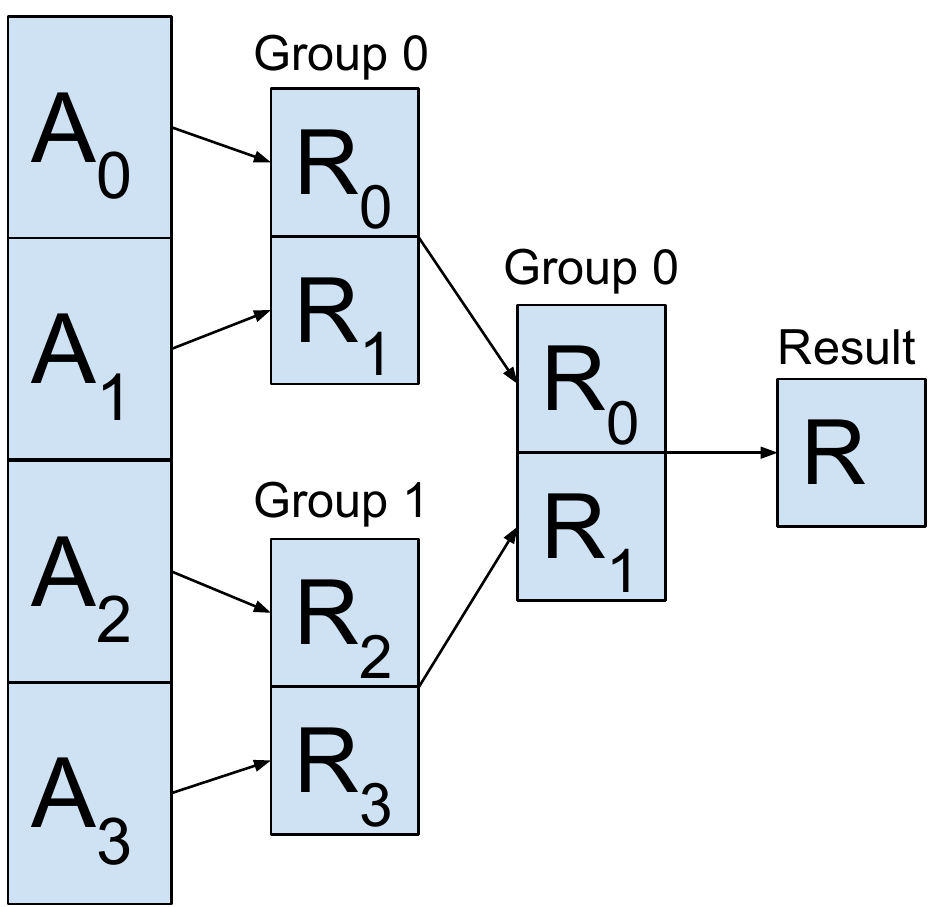}
    \caption{Depiction of the TSQR algorithm \cite{tsqr} used to factor of ``tall skinny" matrices distributed across columns of processors. Each processor first computes a local QR decomposition of its matrix panel $\mathbf{A}_j$. The resulting $\mathbf{R}$ are gathered between processor pairs and then stacked. The process is iterated until each processor contains the same $\mathbf{R}$ factor, which is that of the full $\mathbf{A}$. Only the computation of $\mathbf{R}$ is illustrated; the local $\mathbf{Q}$ factors can be accumulated by having each processor multiply each ``reduced" $\mathbf{Q}$ factor obtained by its successor during each step. }
    \label{fig:TSQRFigure}
\end{figure}

Instead, we use the so-called ``TSQR" algorithm \cite{tsqr}, which computes the reduced QR factorization of a tall and skinny matrix $\mathbf{A}_l$. Tall-and-skinny means that the matrix can be divided into row panels of size $m_r$ such that $m_r \ge N$. Since our $\mathbf{A}_l$ is a slice of $b$ columns from a checkerboard-distributed $\mathbf{A}$, for our purposes this means $M // p_r \ge b$ where $p_r$ is the number of processor rows.

The TSQR algorithm performs the factorization of $\mathbf{A}_l$ via the binary reduction depicted in Figure \ref{fig:TSQRFigure}, with pseudocode given as Algorithm \ref{alg:tsqr}. Each processor in a column computes a local QR decomposition of $\mathbf{A}_l$, yielding a local $\mathbf{R}$ factor. The processors are arranged into groups of two, and the local $\mathbf{R}$ factors gathered within these groups. Pairs of groups are successively combined and the process repeated until only a single $\mathbf{R}$ factor remains, which is that of $\mathbf{A}_l$. 

This procedure yields the reduced factors $\mathbf{Q}_r$ and $\mathbf{R}_r$ of $\mathbf{A}_l$. The full $\mathbf{R}_f$ factor is straightforwardly obtained by appending rows of zeros to $\mathbf{R}_r$. We get the full $\mathbf{Q}_f$ factor implicitly as its so-called $\mathbf{W} \mathbf{Y}$ representation \cite{MatrixComputations}, $\mathbf{Q}_f = \mathbf{I} - \mathbf{W} \mathbf{Y}^H$ where $\mathbf{W}$ and $\mathbf{Y}$ are both $(M, b)$.

To compute $\mathbf{W}$ and $\mathbf{Y}$, we use a slight modification of the ``Yamamoto" procedure outlined in \cite{tsqr_reconstruct}. The Yamamoto procedure has us form 
\begin{subequations}
\label{eq:Yamamoto}
\begin{align}
    \mathbf{Q}_f &= \mathbf{I} - \mathbf{W} \mathbf{T} \mathbf{W}^H \\
    \mathbf{W} &= \mathbf{Q}_r - \mathbf{I} \\
    \mathbf{T}^{-1} &= \mathbf{I} - \mathbf{Q}_1
\end{align}
\end{subequations}
where $\mathbf{T}^{-1}$ is $(b, b)$ and $\mathbf{Q}_1$ is the first $b$ rows of $\mathbf{Q}_r$.  $\mathbf{T}^{-1}$ rather than $\mathbf{T}$ is stored, and multiplications by $\mathbf{T}$ handled via linear solution. This representation is simple to compute, and saves memory compared to the $\mathbf{W}$ $\mathbf{Y}$ form since $\mathbf{T}$ is smaller than $\mathbf{Y}$. Nevertheless, we prefer to form $\mathbf{Y}$ explicitly via $\mathbf{Y}^H = \mathbf{T} \mathbf{W}^H$, so that only one, trivially parallel, linear solve need be performed - compared to one per each multiplication by $\mathbf{Q}_f$.

Note that \eqref{eq:Yamamoto} break down if $\mathbf{T}^{-1}$ is ill-conditioned, which can can occur for example if $\mathbf{Q}_1$ is itself very near to the identity. Said difficulty can be alleviated by a slight generalization described in \cite{tsqr_reconstruct}, replacing each $\mathbf{I}$ in \eqref{eq:Yamamoto} by a diagonal matrix of signs chosen to improve $\mathbf{T}^{-1}$'s conditioning. However, neither \cite{tsqr_reconstruct} nor the references it cites specifies how precisely to choose these. Generalizing from heuristics like ``flip the sign wherever $\mathbf{T}^{-1}$ would otherwise have a row or column of zeros" proves not entirely trivial. Having yet to encounter a practical case of breakdown, we have not implemented the full generalization.

With $\mathbf{W}$ and $\mathbf{Y}$ in hand, we can now straightforwardly perform the \emph{Panel update} step, yielding the full CAQR algorithm. It is depicted in Figure \ref{fig:CAQRFigure} and given as pseudocode in Algorithm \ref{alg:caqr}.

Performance is depicted in the upper right panel of Figure \ref{fig:PerformanceFigure}. For instance, on a full TPU pod (2048 cores) we obtain the QR decomposition of an ($N,N$) matrix of linear size $N = 2^{20} = 1\,048\,576$ in about 20 minutes. Excellent scaling is seen with large-$N$, showing that the task is dominated by the matrix-multiplication update steps. However, our choice of a checkerboard rather than block-cyclic distribution pattern for the matrix $\mathbf{A}$ can result in poor load balancing, since in effect we treat an equally sized matrix at each iteration. Appendix \ref{QRAppendix} shows this to incur about a 3-fold penalty if only $\mathbf{R}$ is computed, or 2.4 if $\mathbf{Q}$ is as well. Note this is at least partially compensated for by the improved single-core throughput in the checkerboard distributed case, achieved by the larger individual blocks fed to the MXUs.

\begin{algorithm}[H]
\caption{TSQR}\label{alg:tsqr}
\begin{algorithmic}[1]
\Require{$(M, N)$ matrix $\mathbf{A}$ distributed among $p_r$ processor rows such that $N > M // p_r$.}
\State group size $\gets 1$.
\State $\mathbf{R}, \mathbf{Q} \gets qr(\mathbf{A})$.
\While{group size $< p_r$}
    \State Double the group size.
    \State Broadcast both unique $\mathbf{R}$ factors within each group.
    \State Vertically stack the newly-broadcast $\mathbf{R}$. 
    \State $\mathbf{Q}, \mathbf{R} \gets qr(\mathbf{R})$.
    \State $\mathbf{Q} \gets \mathbf{Q} \mathbf{Q}_l$ (optional)
\EndWhile
\State \Return $\mathbf{Q}$, $\mathbf{R}$
\end{algorithmic}
\end{algorithm}

\begin{algorithm}[H]
\caption{Right-looking CAQR}\label{alg:caqr}
\begin{algorithmic}[1]
\Require{Checkerboard distributed $(M, N)$ matrix $\mathbf{A}$.}
\State Divide $\mathbf{A}$ into $N_b$ column panels of size $b$, $b = M // N_b$.
\State $\mathbf{Q} \gets \mathbf{I}$.
\For{$j \in [0, N_b)$}
    \State $\mathbf{A}_l \gets $ $(M - j N_b, N_b)$ panel of $\mathbf{A}$ from $\mathbf{A}_{j N_b, j N_b}$.
    \State $\mathbf{A}_r \gets $ all entries in $\mathbf{A}$ right of $\mathbf{A}_l$.
    \State $\mathbf{Q}_r , \mathbf{R}_r \gets$ TSQR$(\mathbf{A}_l)$ (reduced Q factor)
    \State Replace $\mathbf{A}_l$ in $\mathbf{A}$ with $[\mathbf{R}_r, \mathbf{0}]^T$. 
    \State Compute $\mathbf{W}, \mathbf{Y}$ s.t. $\mathbf{Q}_f = \mathbf{I} - \mathbf{W} \mathbf{Y}^H$ (see text)
    \State Replace $\mathbf{A}_r$ in $\mathbf{A}$ with $\mathbf{Q}_f^H \mathbf{A}_r$.
    \State $\mathbf{Q} \gets \mathbf{Q}_{:, M - j N_b:} \mathbf{Q}_f$ (optional).
\EndFor
\State \Return $\mathbf{Q}$, $\mathbf{R} \gets \mathbf{A}$.
\end{algorithmic}
\end{algorithm}

\subsection{Linear solution}
\label{LinearSection}
By ``linear solution" we mean the determination of $\mathbf{x}$ in
\begin{equation}
    \label{eq:LinearSystem}
    \mathbf{A} \mathbf{x} = \mathbf{b}
\end{equation}
where $\mathbf{A}$ and $\mathbf{b}$ are given, with $\mathbf{A}$ an $(N, N)$ matrix, and $\mathbf{x}$ and $\mathbf{b}$ both $(N, k)$. We consider the case of $\mathbf{A}$ given as a dense, full-rank matrix, in which case \eqref{eq:LinearSystem} is typically solved in $O(N^3)$ operations via an initial LU decomposition.

Unfortunately an efficient distributed-TPU LU factorization is not yet available. Instead, we use the QR factorization (as described above), which is more stable and only marginally less efficient. Writing $\mathbf{A} = \mathbf{Q} \mathbf{R}$, we have 
\begin{equation}
    \label{eq:QRSystem}
    \mathbf{R} \mathbf{x} = \mathbf{b}', 
\end{equation}
where $\mathbf{b}' \equiv \mathbf{Q}^H \mathbf{b}$. That is, we have mapped the general linear system in Eq. \eqref{eq:LinearSystem} to the upper triangular one in Eq. \eqref{eq:QRSystem}. In a scalar implementation, such upper triangular systems are trivially soluble by repeated substitution. The row containing a single nonzero element, for example, corresponds to the scalar equation $y_N = x_N$, which is substituted into the row containing two nonzero elements, and so on.

\begin{figure}
    \centering
    \includegraphics[scale=0.5]{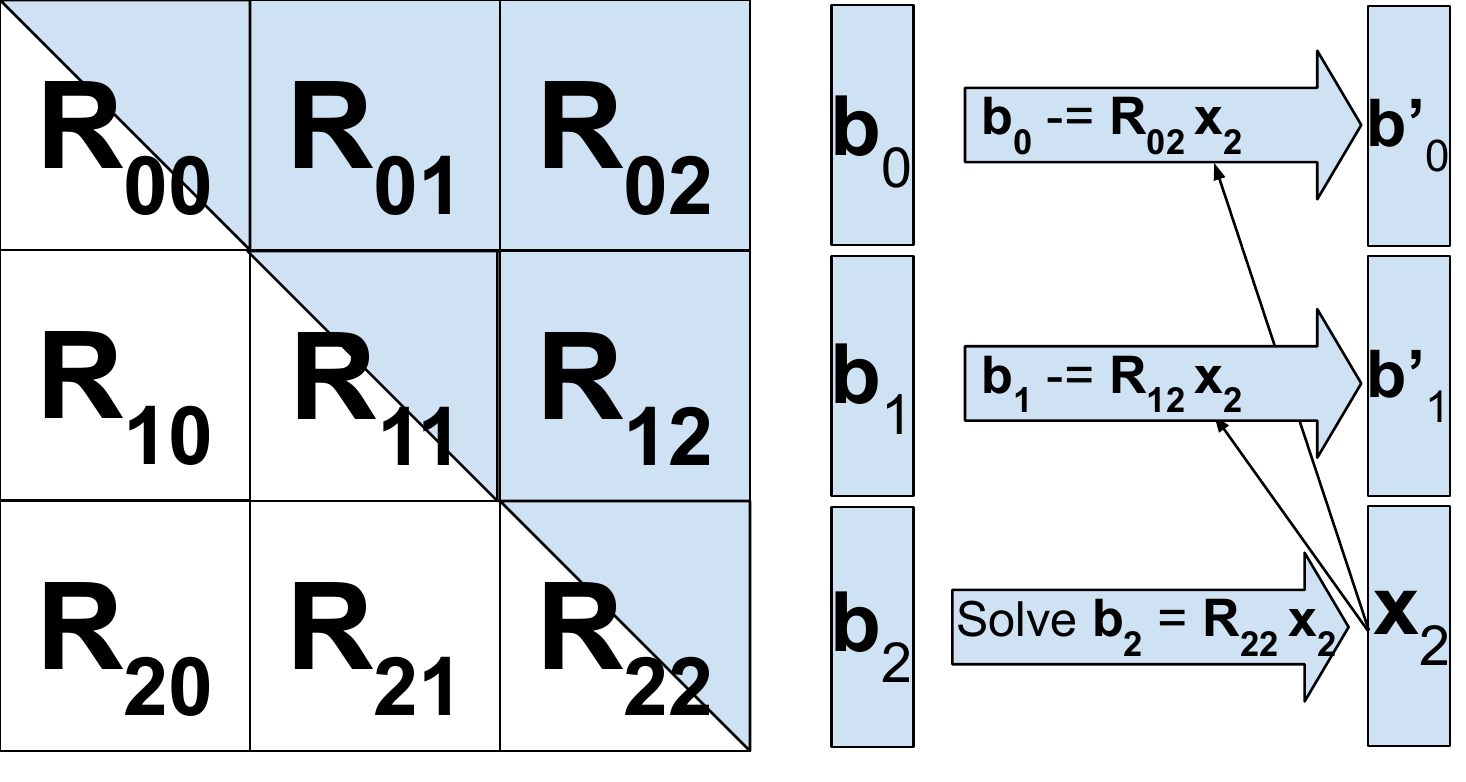}
    \caption{Depiction of our somewhat naiive approach to solving upper triangular systems. We divide the upper triangular matrix $\mathbf{R}$ into square blocks such that each is local to a core and those on the main block diagonal are themselves locally upper triangular. The solution is then found by moving upwards along the main block diagonal from the bottom right - the figure depicts the first such step. At each step, the relevant panel, in this case $\mathbf{x}_2$, of the solution $\mathbf{x}$ is first found by solving the corresponding local triangular system, in this case $\mathbf{R}_{22} \mathbf{x}_2 = \mathbf{b}_2$. The result is broadcast upwards along its column panel, and used to update the coefficients $\mathbf{b}$ as depicted. The procedure then iterates to the upper left.}
    \label{fig:SolveFigure}
\end{figure}

This scalar algorithm is, however, quite TPU unfriendly. Instead, we first note that a reasonably performant single-TPU upper triangular solver, which uses the TPU vector processor and blocking to achieve acceptable performance, ships with Jax. We can leverage this into a naive, but acceptably performant, distributed triangular solver as depicted in Figure \ref{fig:SolveFigure}. The coefficient matrix $\mathbf{R}$ is first divided into square blocks such that each is local to a given processor (a processor may however contain more than one block). The submatrices on the block main diagonal are then themselves upper triangular. 

\begin{algorithm}[H]
\caption{Distributed triangular solver.}\label{alg:solve_triangular}
\begin{algorithmic}[1]
\Require{Checkerboard-distributed $(N, N)$ upper-triangular coefficient matrix $\mathbf{R}$, $(N,)$ RHS vector $\mathbf{b}$ copied between columns.}
\State Divide $\mathbf{R}$ into $N_b$ square blocks of linear size $N // N_b$, indexed with Greek letters.
\State Divide $\mathbf{b}$ into $N_b$ row panels also of size $N // N_b$, indexed with Greek letters.
\For{$\kappa \in (N_b, 0]$}
    \State Solve $\mathbf{R}_{\kappa \kappa} \; \mathbf{x}_\kappa = \mathbf{b}_\kappa$.
    \State Broadcast $\mathbf{x}_\kappa$ upwards along its processor column.
    \State $\mathbf{b}_\kappa \gets \mathbf{x}_\kappa$
    \For{$\gamma \in [0, \kappa)$}  \Comment Do this in parallel.
        \State $\mathbf{b}_\gamma \gets \mathbf{b}_\gamma - \mathbf{A}_{\gamma \kappa} \mathbf{x}_\kappa$.
    \EndFor
    \State Copy the updated $\mathbf{b}$ to the other processor columns.
\EndFor
\State \Return $\mathbf{b}$
\end{algorithmic}
\end{algorithm}
From here, we perform a direct blocked analogy of the scalar elimination procedure described above, with each column panel of $\mathbf{R}$ treated in serial. First, the triangular system at the bottom of the panel is solved (e.g. $\mathbf{R}_{22} \mathbf{x}_2 = \mathbf{b}_2$ in Figure \ref{fig:SolveFigure}). The resulting $\mathbf{x}_i$ is the panel of the full solution overlapping its corresponding $\mathbf{b}_i$, and if desired $\mathbf{b}_i$ may be overwritten by it in place. The corresponding substitution is achieved by broadcasting $\mathbf{x}_i$ to the blocks above it, and subtracting $\mathbf{R} \mathbf{x}_i$ from each $\mathbf{b}$ panel above. Pseudocode is given as Algorithm \ref{alg:solve_triangular}.

After the initial QR factorization, this algorithm is poorly load-balanced. The cores storing zeroes of $\mathbf{R}$ are left completely idle; only the update step runs in parallel; and during it only the cores above the current main block diagonal do work. Much better load balancing could be achieved by adopting a block cyclic data distribution, so that the processor grid was not so tightly coupled to the matrix block locations. The algorithm is, however, sufficiently efficient to represent a small expense compared to the QR step, as can be seen in the bottom-left panel of Figure \ref{fig:PerformanceFigure}. As an example, for an ($N,N$) matrix $\mathbf{A}$ of linear size $N = 2^{20} = 1\,048\,576$, we can solve a linear system on a full TPU pod (2048 cores) again in about 20 minutes.

\subsection{Matrix Functions}
Above we considered application of TPU slices towards bread and butter tasks in scientific computing. 
Since TPUs are natively optimized for matrix multiplication, it is most natural to consider also tasks based on matrix multiplication, such as matrix functions (implemented approximately as matrix polynomials, thus requiring matrix multiplications and additions). Next we briefly review two types of matrix functions that transform, respectively, the singular values and the eigenvalues of a matrix.

Recall first that every matrix has a \textbf{singular value decomposition} (SVD),
\begin{equation}
    \label{eq:SVD}
    \mathbf{A} = \mathbf{U}_s \, \Sigma \, \mathbf{V}^H_s,
\end{equation}
with $\Sigma$ a diagonal matrix of real \textbf{singular values} and $\mathbf{U}_s$ and $\mathbf{V}^H_s$, the left and right \textbf{singular vectors}, both unitary. Given any polynomial $f(x)=a_1x + a_3x^3 + a_5x^5+\cdots$ made only of odd powers of $x$, we can define the matrix function $f(\mathbf{A})$ acting on the singular values of $\mathbf{A}$,
\begin{equation}
    \label{eq:MatFunc}
    f(\mathbf{A}) \equiv \mathbf{U}_s \, f(\Sigma) \, \mathbf{V}^H_s~~~~~\mbox{(singular values)},
\end{equation}
where $f(\Sigma)$ is a diagonal matrix where each diagonal entry contains the result of applying $f$ to the corresponding singular value in $\Sigma$. Notice that $\mathbf{A}$ and $\mathbf{A} (\mathbf{A}^{\dagger} \mathbf{A})^n$ share the same structure of singular vectors for any integer $n=0,1,2,\cdots $, that is
\begin{equation} \label{eq:odd_power}
    \mathbf{A}(\mathbf{A}^{\dagger}\mathbf{A})^n = \mathbf{U}_s \, \Sigma^{2n+1} \, \mathbf{V}^H_s.
\end{equation}
It then follows that we can compute $f(\mathbf{A})$ by means of the matrix polynomial expansion
\begin{eqnarray} \label{eq:OddPoly}
    &&a_1\mathbf{A} + a_3 \mathbf{A}\mathbf{A}^H \mathbf{A} + a_5 \mathbf{A}\left(\mathbf{A}^H \mathbf{A}\right)^2 + \ldots\\
    &=& \mathbf{U}_s \left( a_1\Sigma + a_3\Sigma^3 + a_5\Sigma^5 +\cdots \right) \mathbf{V}^H_s\\
    &=& \mathbf{U}_s \, f(\Sigma) \, \mathbf{V}^H_s = f(\mathbf{A}).
\end{eqnarray}

Recall now that every diagonalizable square matrix also has an \textbf{eigenvalue decomposition} (EVD),
\begin{equation}
    \label{eq:EVD}
    \mathbf{A} = \mathbf{P}_e \, \Omega \, \mathbf{P}^{-1}_e,
\end{equation}
with $\Omega$ a diagonal matrix of (possibly complex) \textbf{eigenvalues} and $\mathbf{P}_e$ an invertible matrix whose columns encode the right \textbf{eigenvectors} of $\mathbf{A}$. Given an arbitrary polynomial $g(x) = a_0 + a_1 x + a_2 x^2 + \cdots$, we can define the matrix function $g(\mathbf{A})$ for a diagonalizable square matrix $\mathbf{A}$ by acting on its eigenvalues,
\begin{equation}
    \label{eq:MatFuncEig}
    g(\mathbf{A}) \equiv \mathbf{P}_e \, g(\Omega) \, \mathbf{P}^{-1}_e~~~~~\mbox{(eigenvalues)},
\end{equation}
with $g(\Omega)$ a diagonal matrix where each diagonal entry contains the result of applying $g$ to the corresponding eigenvalue in $\Omega$. We emphasize that this definition of matrix function, based on transforming the eigenvalues while preserving the structure of eigenvectors, is not equivalent to that in Eq. \eqref{eq:MatFunc}, which transformed the singular values while preserving the singular vectors. We observe that the matrices $\mathbf{A}^n$ for $n=0,1,2,\cdots$ share the same structure of eigenvectors, that is
\begin{equation} \label{eq:any_power}
    \mathbf{A}^n = \mathbf{P}_e \, \Omega^n \, \mathbf{P}^{-1}_e.
\end{equation}
It then follows that we can compute $g(\mathbf{A})$ by means of the matrix polynomial expansion
\begin{eqnarray}
\label{eq:EigenvaluePolynomial}
&&a_0 \mathbf{I} + a_1 \mathbf{A}+ a_2 \mathbf{A}^2 + \cdots \\
&=& \mathbf{P}_e \left(a_0\mathbf{I} + a_1\Omega + a_2\Omega^2 + \cdots \right) \mathbf{P}^{-1}_e\\
&=& \mathbf{P}_e g(\Omega) \mathbf{P}^{-1}_e = g(\mathbf{A}). 
\end{eqnarray}



Various matrix functions of interest, such as matrix sign function and matrix inverse (see below), but also matrix principal square root, matrix inverse principal square root, matrix exponential, matrix logarithm, etc, can be accurately approximated by polynomials (or polynomial iterations) of one of the two forms above, and thus efficiently computed and scaled on TPUs. Here we illustrate this with the so-called $\textbf{polar decomposition}$, which is obtained through applying the sign function to the singular values, where the sign function is approximated by means of a polynomial iteration made of small polynomials of the type in Eq. \eqref{eq:OddPoly}.

The polar decomposition of an arbitrary $(M,N)$ matrix $\mathbf{A}$ with $M\geq N$ is defined by
\begin{equation}
    \label{eq:PolarDecomp}
    \mathbf{A} = \mathbf{U} \mathbf{H},
\end{equation}
where the $(M,N)$ matrix $\mathbf{U}$ has $N$ orthonormal columns and the $(N,N)$ matrix $\mathbf{H}$ is positive semi-definite. This is a matrix version of the polar decomposition $z=e^{i\phi} |z|$ of a complex number $z$ into its complex phase $e^{i\phi}$ and its non-negative norm $|z|$.
In terms of the SVD \eqref{eq:SVD} we have
\begin{equation}
    \label{eq:SVDdef}
    \mathbf{U} = \mathbf{U}_s \mathbf{V}^H_s,
\end{equation}
i.e. that the polar factor $\mathbf{U}$ can be obtained by setting all the singular values of $\mathbf{A}$ to $1$ while leaving the singular vectors untouched. 

It is easy to confirm that repeated application of the scalar polynomial iteration
\begin{equation}
    \label{eq:NS_scalar}
    x_{i+1} = \frac{1}{2}x_i (3 - x_i^2)    
\end{equation}
sends any initial $x_0 \in (0, \sqrt{3}) \to +1$, while sending $x_0 = 0$ to $0$. In other words, this polynomial iteration converges to the \textbf{sign} function when applied on the interval $[0,\sqrt{3})$. The corresponding matrix polynomial, the \textbf{Newton-Schulz iteration},
\begin{equation}
    \label{eq:NewtonSchulz}
    \mathbf{X}_{i+1} = \frac{1}{2}\mathbf{X}_i(3\mathbf{I} - \mathbf{X}_i^H \mathbf{X}_i),
\end{equation}
starting with $\mathbf{X}_0 \equiv \mathbf{A}$, thus has the same effect upon the singular values of $\mathbf{A}$, and therefore has the unitary polar factor of $\mathbf{U}$ in Eq. \eqref{eq:SVDdef} as its fixed point. 
As confirmed in \cite{NakatsukasaHighamStable}, this iteration is numerically stable for any $\mathbf{A}$ with $||\mathbf{A}||_2 < \sqrt{3}$, where $||\mathbf{A}||_2$ denotes the spectral 2-norm of $\mathbf{A}$, or its largest singular value. We can ensure this property for general input by an initial rescaling. We use $\mathbf{A} \to \mathbf{X}_0 \equiv \frac{\mathbf{A} (\sqrt{3} - \delta)}{||\mathbf{A}||_F}$, where $||\mathbf{A}||_F$ denotes the Frobenius norm (which can be computed easily as $\sqrt{\mbox{tr} (\mathbf{A}\mathbf{A}^H)}$ and fulfils $||\mathbf{A}||_F \geq ||\mathbf{A}||_2$ ) and $\delta$ is an arbitrary, small positive number.

Once the smallest singular value $s_0$ of $\mathbf{A}$ grows to $0.1$ or so, through the Newton-Schulz iterations \eqref{eq:NewtonSchulz}, it then subsequently enjoys quadratic convergence to 1. When working in single precision, this means that it only requires about 10 further iterations before it reaches $1$ within that precision. Convergence before this point can unfortunately be rather slow, so that 35-50 iterations might be required if $s_0$ is initially very small. 

\begin{figure}
    \label{fig:NS_scalars}
    \centering
    \includegraphics[scale=0.3]{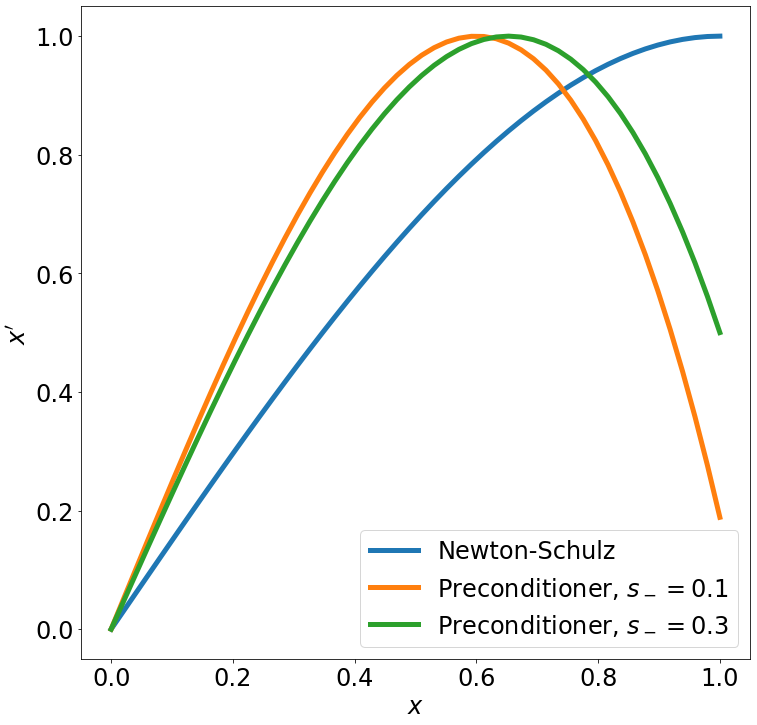}
    \caption{The Newton-Schulz polynomial \eqref{eq:NS_scalar}, alongside the preconditioning polynomial \eqref{eq:RogueScalar} for different choices of $s_-$. No input is mapped by \eqref{eq:RogueScalar} beneath $s_-$, but compared to \eqref{eq:NS_scalar}, the slope of the latter is much larger near 0.}
    \label{fig:my_label}
\end{figure}

To improve upon this, we choose a desired minimum singular value $s_-$, and apply a preconditioning polynomial
\begin{subequations}
\begin{align}
    \label{eq:a_def}
    a &= \frac{3}{2} \sqrt{3} - s_- \\
    \label{eq:RogueScalar}
    x_{i+1} &= a x_i\left(1 - \frac{4}{27} (a x_i)^2\right) \\
    \label{eq:RogueMatrix}
    \mathbf{X}_{i+1} &= a \mathbf{X}_i\left(\mathbf{I} - \frac{4}{27} a^2 \mathbf{X}_i^H \mathbf{X}_i\right)
\end{align}
\end{subequations}
While \eqref{eq:RogueScalar} does not monotonically drive values towards 1  (see Figure \ref{fig:NS_scalars}), it does monotonically drive any beneath $s_-$ upward, more quickly than \eqref{eq:NS_scalar}, while keeping those larger comfortably above the threshold $[s_-, 1]$. Consequently, \eqref{eq:RogueMatrix} rapidly improves the conditioning of $\mathbf{X}$ without affecting its singular vectors. The number of applications needed to obtain a spectrum in $[s_-, 1]$ can be tracked by repeatedly feeding an estimated initial minimum singular value $s_0$ through \eqref{eq:RogueScalar} until a value greater than $s_-$ is obtained. We typically use $s_0 = \epsilon$, the machine precision - about $10^{-7}$ in single precision - which requires about 10-15 iterations for $s_- = 0.1$. Notice that in order to use \eqref{eq:a_def}-\eqref{eq:RogueMatrix}, we want to rescale the initial matrix $\mathbf{A}$ to have singular values in the interval $[0,1)$, which we achieve through $\mathbf{A} \to \mathbf{X}_0 \equiv \frac{\mathbf{A} (1 - \delta)}{||\mathbf{A}||_F}$ for some small $\delta > 0$.

Algorithm \ref{alg:NS_Polar} summarizes our approach. In total, it then takes about 25 iterations to obtain the polar factor of an arbitrary matrix, which could potentially be reduced to about 10 for well-conditioned input with $s_0 = 0.1$. Thus, this operation is equivalent to about 50 matrix multiplications. This is demonstrated in the bottom-right of Figure \ref{fig:PerformanceFigure}, essentially a rescaling of the top-left panel by a factor of about 50. Memory footprint, scaling, and use of hardware resources follow the same reasoning as for SUMMA, since the algorithm consists simply of repeated calls to SUMMA. As an example, on a full TPU pod (2048 cores) we can compute the polar decomposition of an $(N,N)$ matrix of linear size $N = 2^{19}= 524\, 288$ in about 20 minutes.

As alluded to above, various iterations besides that leading to the polar decomposition can also be efficiently implemented. For example, for the electronic structure DFT computations presented in \cite{tpu_qchem}, the matrix inverse square root of an overlap matrix for single-electron basis functions needs to be computed, as well as a so-called purification of a Hermitian matrix that is similar to the polar decomposition described above. The Newton-Schulz procedure may also be used to compute matrix inverses, as detailed in Algorithm \ref{alg:NS_Inverse}. 

\begin{algorithm}[H]
\caption{Preconditioned Newton-Schulz Polar Factorization}\label{alg:NS_Polar}
\begin{algorithmic}[1]
\Require{$(M, N)$ matrix $\mathbf{A}$ with $M \ge N$, threshold $s_- \sim 0.1$, error tolerance $\epsilon$, estimated smallest singular value $s_0$.}
\Ensure{Isometric $\mathbf{U}$ giving positive semi-definite $\mathbf{H} \equiv \mathbf{U}^H \mathbf{A}$.}

\State $\mathbf{U} \gets \mathbf{A} / ||\mathbf{A}||_F $.
\State $s \gets s_0$ \Comment Or $\epsilon$ if unsupplied.
\State $a \gets \frac{3}{2} \sqrt{3} - s_-$
\While{$s < s_-$} \Comment Bounds singular values by $[s_-, 1]$.
    \State $s \gets a s(1 - \frac{4}{27}a^2 s^2)$
    \State $\mathbf{U} \gets a \mathbf{U}(\mathbf{I} - \frac{4}{27} a^2 \mathbf{U}^H \mathbf{U})$
\EndWhile

\While{$\delta > \mathrm{max}(M, N) \epsilon$}
    \State $\mathbf{U}' \gets \frac{1}{2}\mathbf{U}(3\mathbf{I} - \mathbf{U}^H \mathbf{U})$
    \State $\delta \gets ||\mathbf{U}' - \mathbf{U}||_F$
    \State $\mathbf{U} \gets \mathbf{U}'$
\EndWhile
\State \Return $\mathbf{U}$
\end{algorithmic}
\end{algorithm}

\begin{algorithm}[H]
\caption{Newton-Schulz Matrix Inversion}\label{alg:NS_Inverse}
\begin{algorithmic}[1]
\Require{$(N, M)$ full-rank matrix $\mathbf{A}$.}
\Ensure{$\mathbf{X}$ s.t. one of $\mathbf{X} \mathbf{A} = \mathbf{I}$ or $\mathbf{A} \mathbf{X} = \mathbf{I}$.}

\State $a \gets  ||\mathbf{A} \mathbf{A}^H||_F$.
\State $\mathbf{X} \gets \frac{1}{a} \mathbf{A}^H $.
\State $\epsilon \gets \mathbf{u}$ \Comment $\mathbf{u}$ is unit roundoff.
\State $\delta \gets 2 \epsilon$ \Comment Initial error g.t. tolerance.
\While{$\delta > \epsilon$}
    \If{$\mathbf{X} \mathbf{A} = \mathbf{I}$ desired}
        \State $\mathbf{X}' \gets \mathbf{X} (2\mathbf{I} - \mathbf{A}\mathbf{X})$ 
    \ElsIf{$\mathbf{A} \mathbf{X} = \mathbf{I}$ desired}
        \State $\mathbf{X}' \gets (2\mathbf{I} - \mathbf{AX})\mathbf{X}$
    \EndIf
    \State $\delta \gets ||\mathbf{X}' - \mathbf{X}||_F$
    \State $\epsilon \gets 2 \epsilon$  \Comment Note growing error tolerance.
    \State $\mathbf{X} \gets \mathbf{X}'$
\EndWhile
\State \Return $\mathbf{X}$
\end{algorithmic}
\end{algorithm}

We can in fact approximate any sufficiently smooth function with a polynomial expansion.
Given a scalar function $g$, if we know that the eigenvalues of a square matrix $\mathbf{A}$ are within some interval $[a, b]$, and we have a polynomial $p_d$ that approximates $g$ to a desired accuracy within that interval, we can evaluate $p_d(\mathbf{A})$ as in~\eqref{eq:EigenvaluePolynomial} to approximate $g(\mathbf{A})$ in the sense of~\eqref{eq:MatFuncEig}.
Naive polynomial expansions are often oscillatory, but expansions in terms of Chebyshev polynomials minimise such oscillations, making them ideal for this use.
The accuracy of the approximation is controlled by the degree $d$ of the polynomial $p_d$, and evaluating $p_d(\mathbf{A})$ requires $d$ matrix products, using the so called Clenshaw summation method~\cite{gil2007numerical}.
How large a $d$ is needed for a given accuracy depends on both the smoothness of $g$ and the spectrum of $\mathbf{A}$, but the advantage of this method is that it can be easily applied to any piece-wise smooth $g$.

\section{Conclusion}
\label{Conclusion}

In this paper we have demonstrated the potential of TPUs to serve as accelerators for large-scale scientific computation, by using distributed, matrix-multiply-based algorithms for the QR decomposition, solving a linear system, and matrix functions such as the polar decomposition (see also Appendix \ref{Appendix2}). By distributing the matrices over a full pod of third generation TPUs (2048 cores), large matrices with linear size up to $N=2^{20}= 1\, 048\, 576$ can be addressed, with computational times ranging from 2 minutes (for matrix multiplication) to 20 minutes (e.g. for QR decomposition). Moreover, a full pod of fourth generation TPUs (8192 cores) is expected to address matrices that double the above linear size in comparable times (work in progress). As shown in subsequent papers, see \cite{tpu_circuit, tpu_Z2field, tpu_qphys, tpu_floquet, tpu_qhardware, tpu_qchem, tpu_CC, tpu_DMRG}, the technology demonstrated here is already significant for a wide range of applications in the context of large-scale simulations and computations of quantum systems, including quantum computation, quantum many-body physics, quantum chemistry, and materials science.

Machine learning ASICs are broadly accessible as a cloud service. For instance, anyone with a Google Cloud Platform account can have access to a TPU pod. As a result, a number of large-scale scientific computing tasks such as the ones demonstrated in this paper and in \cite{tpu_circuit, tpu_Z2field, tpu_qphys, tpu_floquet, tpu_qhardware, tpu_qchem, tpu_CC, tpu_DMRG} are now within reach of any reach group, contributing to democratizing supercomputing throughout the scientific community and beyond.

\acknowledgements
This work would not have been possible without the at-times-heroic support from the Google teams associated with Jax and with Cloud TPUs, including but not limited to Skye Wanderman-Milne, Rasmus Larsen, Peter Hawkins, Adam Paszke, Stephan Hoyer, Sameer Agarwal, Matthew Johnson, Zak Stone, and James Bradbury. 
The authors also thank Chase Riley Roberts, Jae Yoo, Megan Durney, Stefan Leichenauer and the entire Sandbox@Alphabet team for early work, discussions, encouragement and infrastructure support. 
This research was supported with Cloud TPUs from Google's TPU Research Cloud (TRC). Sandbox is a team within the Alphabet family of companies, which includes Google, Verily, Waymo, X,  and others. GV is a CIFAR fellow in the Quantum Information Science Program and a Distinguished Visiting Research Chair at Perimeter Institute. Research at Perimeter Institute is supported by the Government of Canada through the Department of Innovation, Science and Economic Development and by the Province of Ontario through the Ministry of Research, Innovation and Science.

\bibliography{references}

\clearpage

\appendix

\section{Complexity penalty due to checkerboard-distributed QR}
\label{QRAppendix}
In this Appendix we compute the complexity penalty incurred by our choice of a checkerboard rather than block-cyclic distribution in our distributed QR algorithm. Due to our adoption of a checkerboard rather than a block-cyclic distribution, successive iterations of this algorithm must operate on the full matrix at each step, rather than the shrinking submatrix which logically need be treated.

More quantitatively, let us consider the case that $\mathbf{Q}$ is not computed, so that the dominant expense of the algorithm is the update of $\mathbf{A}$ (line 7 in Algorithm \ref{alg:caqr}). Let $\tilde{M}$ and $\tilde{N}$ be the respective dimensions of the matrix block being updated. The total cost $c$ of the updates is then
\begin{equation}
    c = \sum_{i=0}^{i=\frac{N}{b}-1} 4b\tilde{M} \tilde{N}.
\end{equation}
Suppose one could correctly reduce the size of the updated block during the computation, as would be made possible by a block-cyclic data distribution. At iteration $i$ we have $\tilde{M} = M - (i + 1)b$ and $\tilde{N} = N - (i + 1)b$, and thus a block-cyclic expense $c_{\mathrm{bc}}$ of
\begin{equation}
    c_{\mathrm{bc}} = 2MN^2 - \frac{2}{3}N^3.
\end{equation}
However, using our checkerboard distribution, processors must work as if the block does not reduce in size. We then have $\tilde{M} = M$, $\tilde{N} = N$, and thus a checkerboard-distributed expense $c_{\mathrm{chk}}$ of
\begin{equation}
    c_{\mathrm{chk}} = 4MN^2.
\end{equation}
The ratio of these factors, which is 3 in the $N = M$ case (decreasing to 2.4 if $\mathbf{Q}$ is also computed), is the unrealized optimization offered by a block-cyclic distribution. The optimization could be realized fairly straightforwardly --- though with some risk of losing single-core throughput speed due to the smaller local matrix sizes --- but we leave doing so to future studies.

\section{Further optimizations and other experiments}
\label{Appendix2}

As mentioned in the text significant optimization opportunity remains. Efficient distributed Cholesky and LU implementations would also be of great interest. We are currently developing a symmetric eigensolver based on a two-sided ``band reduction" variation of the QR solver.

It is interesting to briefly detail approaches which have not been so successful. First, following the ideas in \cite{HighamMixed}, we at one point attempted to use Algorithm \ref{alg:NS_Inverse} to compute a low-precision inverse with which to precondition a GMRES-based linear solver. While this does work, in the end the QR approach is simply too much more efficient for this to be useful. Second, following a ``spectral divide and conquer" approach described in \cite{NakatsukasaHigham2013}, either the polar factorization or the above purification routine may be successively applied to compute progressively smaller submatrices containing only half of an input matrix's eigenvalue spectrum, theoretically leading to an efficient Hermitian eigensolver based only on matrix multiplication. We have implemented such an eigensolver, but have found it to be lacking in both stability and efficiency in practice, due partly to XLA's need to recompile upon encountering matrices of new size. 

\end{document}